\documentclass[aps,twocolumn,showpacs,superscriptaddress]{revtex4-1}
\usepackage{graphicx,enumerate}
\usepackage{overpic}
\usepackage{amsmath,amsfonts}
\usepackage{bm}
\usepackage[colorlinks,,urlcolor=blue,linkcolor=blue,anchorcolor=blue,citecolor=blue]{hyperref}
\def\d{\mathrm{d}}
\def\sgn{\mathrm{sgn}}

\begin{document}
\title{Towards large-Chern-number topological phases by periodic quenching}
\author{Tian-Shi Xiong}
\affiliation{Center for Interdisciplinary Studies $\&$ Key Laboratory for Magnetism and Magnetic Materials of the MoE, Lanzhou University, Lanzhou 730000, China}
\author{Jiangbin Gong}\email{phygj@nus.edu.sg}
\affiliation{Department of Physics, National University of Singapore, Singapore 117543}
\author{Jun-Hong An}\email{anjhong@lzu.edu.cn}
\affiliation{Center for Interdisciplinary Studies $\&$ Key Laboratory for Magnetism and Magnetic Materials of the MoE, Lanzhou University, Lanzhou 730000, China}

\begin{abstract}
  Topological phases with large Chern numbers have important implications. They were previously predicted to exist by considering fabricated long-range interactions or multi-layered materials. Stimulated by recent wide interests in Floquet topological phases, here we propose a scheme to engineer large-Chern-number phases with ease by periodic quenching. Using a two-band system as an example, we theoretically show how a variety of topological phases with widely tunable Chern numbers can be generated by periodic quenching between two simple Hamiltonians that otherwise give low Chern numbers.  The obtained large Chern numbers are explained through the emergence of multiple Dirac cones in the Floquet spectra.  The transition lines between different topological phases in the two-band model are also explicitly found, thus establishing a class of easily solvable but very rich systems useful for further understandings and applications of topological phases in periodically driven systems.
\end{abstract}
\date{\today}
\maketitle

\section{Introduction}\label{Introduction}
Topological insulators behave as an insulator in their interior but contain topologically protected conducting surface states. Their importance lies in not only electronic systems but also photonic analog systems \cite{RevModPhys.82.3045, RevModPhys.83.1057, RevModPhys.80.1083, Nature.08293, PhysRevLett.100.013904}. The robustness of the surface states to weak disorder may be traced back to the topological invariants
of the interior bulk bands. For two-dimensional systems, the band topology can be characterized by the Chern numbers. In particular, large-Chern-number topological phases \cite{PhysRevLett.111.136801, PhysRevLett.112.046801, PhysRevLett.113.113904, PhysRevB.85.045445, PhysRevB.86.241111, PhysRevX.2.031013, PhysRevLett.115.126401, PhysRevLett.115.087003, PhysRevLett.114.236803, PhysRevLett.115.253901, PhysRevB.93.094521} have attracted considerable interests.  Providing more edge channels, these intriguing phases are expected to improve certain device performance by lowering the contact resistance in quantum anomalous hall insulators \cite{PhysRevLett.112.046801, PhysRevLett.111.136801}.  In photonic crystals, a phase with a large Chern number can support multi-mode one-way waveguides, a feature relevant to realizing  reflectionless waveguides splitters, combiners, and one-way photonic circuits \cite{PhysRevLett.100.013904, PhysRevLett.113.113904, PhysRevLett.115.253901}.

Natural materials, even topologically nontrivial, normally do not (unless there is long-range hopping) produce a band structure with large Chern numbers \cite{PhysRevB.85.165456, PhysRevB.87.115402}.
Indeed,  the Chern numbers observed in recent experiments were mostly $\pm1$ \cite{Chang12042013, PhysRevLett.111.136801, PhysRevLett.112.046801, PhysRevLett.113.113904}. Large-Chern-number phases are expected to emerge if multiple-layer structures are fabricated \citep{PhysRevX.2.031013, PhysRevB.85.045445, PhysRevB.86.241111, PhysRevB.86.241112}. In the context of photonic crystals, topological phases with Chern numbers up to 4 were obtained by increasing the number of band touching points via proper material fabrication \cite{PhysRevLett.113.113904, PhysRevLett.115.253901}.

We take an alternative route, namely, the generation of topologically nontrivial Floquet states \cite{PhysRevB.79.081406, PhysRevB.82.235114, Nphys.1926, PhysRevLett.108.056602, PhysRevLett.110.016802, PhysRevLett.109.010601, PhysRevB.90.195419,PhysRevB.90.205108,PhysRevLett.110.200403}, to realize large-Chern-number phases in driven systems. The flexibility and versatility of periodic driving in generating Floquet topological phases can be easily recognized. Instead of having band features once a material is fabricated, the Floquet bands of a periodically driven system (topology, band width etc.) may be engineered on demand by tuning the driving-field parameters. The idea of Floquet topological states has been tested in many systems including graphene \cite{PhysRevLett.113.236803, PhysRevB.89.121401, PhysRevB.90.115423}, optical lattices \cite{PhysRevX.4.031027, PhysRevA.89.063628, PhysRevA.84.053601, PhysRevA.91.033601, PhysRevE.91.052129, PhysRevA.91.020101, Ncomms.9336}, and photonic crystals \cite{Nature.12066}, also leading to the discovery of novel topological phases totally absent in static systems \cite{Wang25102013, PhysRevLett.113.266801, PhysRevX.3.031005}. In addition, as argued in a recent study by two of us and collaborators \cite{PhysRevB.87.201109}, from the perspective of the one-period effective Hamiltonian of a periodically driven system,  periodic driving can effectively synthesize long-range interactions and manipulate the time-reversal symmetry, two factors relevant to the realization of large-Chern-number phases.

Specifically, we exploit a periodic quenching protocol applied to two-dimensional systems. That is, the overall system Hamiltonian is periodically switched between two simple ones. Consider, for example, the Haldane model, one version of which has been experimentally realized \cite{Nature.13915}. Then such a protocol may be realized by periodically quenching certain system parameters using a similar experimental setup. Using two-band systems, we obtain large-Chern-number phases and explicitly find the theoretical boundaries between different topological phases. For a periodically quenched (generalized) Haldane model, the obtained Chern numbers are found to range from $-7$ to $+7$ and explained via the generation of multiple Dirac cones in the Floquet spectra.

Our paper is organized as follows. In Sec. \ref{Two-band systems}, we show the periodically quenched two-band systems and derive the analytical results on their Floquet topological phase transition. In Sec. \ref{Results}, our analytical expectation is explicitly confirmed in the generalized Haldane model and the topological phases with large Chern number are found. Finally, a summary is given in Sec. \ref{Conclusions}.

\section{Periodically quenched two-band systems}\label{Two-band systems}
Consider a general static two-dimensional two-band momentum-space Hamiltonian as follows \cite{PrincetonUP.2013},
\begin{equation}\label{HPauli}
\mathcal{H}(\mathbf{k})=\varepsilon(\mathbf{k})I_{2\times2}+\mathbf{h}(\mathbf{k})\cdot{\pmb \sigma},
\end{equation}
where ${\pmb\sigma}$ are the Pauli matrices, $I_{2\times2}$ is the identity matrix, and ${\bf k}=(k_{x}, k_{y})$ is the two-dimensional momentum vector. The nontrivial topological property of this system is evidenced by the band Chern number, which can be calculated (for the lower band) from \cite{PhysRevLett.49.405}
\begin{equation}\label{MTChern}
\mathcal{C}=\frac{1}{4\pi}\int_{\text{BZ}}\d^{2}\mathbf{k}\ \frac{\mathbf{h}}{|\mathbf{h}|^{3}}\cdot(\partial_{k_{x}}\mathbf{h}\times\partial_{k_{y}}\mathbf{h}),
\end{equation}
where BZ stands for the Brillouin zone. In the following, certain parameters of ${\mathbf h}$ will be periodically quenched between two different sets of values.  We shall hence use different, but analogous topological invariants to characterize the topology of periodically driven systems.

We start from general discussions on the Floquet states of a time-periodic Hamiltonian with $ H(t)=H(t+T)$. The eigensolution of the one-period evolution operator $U_T=\mathcal{T}\exp[-i\int_0^T  H( t)dt]$ fully determines the dynamics at multiples of $T$.  Let $U_T|u_\alpha\rangle=e^{-i\epsilon_\alpha T}|u_\alpha\rangle$, where $\epsilon_\alpha$ are the quasienergies and $|u_\alpha\rangle$ are the Floquet eigenstates. The quasienergies $\epsilon_\alpha$, defined up to a multiple of $2\pi/T$ and generally chosen to lie in $(-\pi/T,\pi/T]$, form the Floquet bands. When the translational symmetry is maintained in $H(t)$, we can solve the eigenequation in momentum space to obtain $|u_\alpha(\mathbf{k})\rangle$. A topological characterization of the Floquet bands can be done by obtaining the Chern numbers of the Floquet states $|u_\alpha({\bf k})\rangle$. Because all the Berry curvature information is fully contained in $|u_\alpha(\mathbf{k})\rangle$, the calculation of the Floquet-band Chern numbers can be assisted by defining the one-period effective Hamiltonian ${\cal H}_{\text{eff}}({\mathbf k})\equiv \frac{i}{T}\ln\left[U_T({\mathbf k})\right]$, which shares precisely the same eigenstates as $U_T({\mathbf k})$.  That is, it is convenient to obtain ${\cal H}_{\text{eff}}({\mathbf k})$ in the form of Eq.~(\ref{HPauli}) and then calculate the Floquet-band Chern numbers using again Eq.~(\ref{MTChern}).

Unlike other forms of periodic driving (such as continuous harmonic driving \cite{PhysRevB.93.144307, EPJB.87204}), periodic quenching \cite{PhysRevB.87.201109} allows for analytical studies of two-band systems. Under a periodic quenching protocol of period $T=T_1+T_2$, we assume the system Hamiltonian $H_1$ ($H_2$) possesses the momentum-space Hamiltonian ${\cal H}_{1}$ (${\cal H}_2$) for $T_1$ ($T_2$). Then ${\cal H}_{\text{eff}}({\mathbf k})$ is given by
\begin{equation}\label{Heff}
\mathcal{H}_\text{eff}(\mathbf{k})\equiv\frac{i}{T_{1}+T_{2}}\ln [e^{-i\mathcal{H}_{2}(\mathbf{k})T_{2}}e^{-i\mathcal{H}_{1}(\mathbf{k})T_1}].
\end{equation}
Note that $\mathcal{H}_\text{eff}$ defined above as well as the associated Floquet-band Chern numbers can be explicitly obtained  by use of the Pauli-matrix algebra.

A number of results are in order: (i) It can be derived that the two Floquet bands of $U_T({\mathbf k})$ would touch each other when the following two conditions are met:
\begin{align}
&\hat{\mathbf{h}}_{1}(\mathbf{k})=\pm\hat{\mathbf{h}}_{2}(\mathbf{k}),\tag{4.a}\label{Con.a}\\
&T_{1}|\mathbf{h}_{1}(\mathbf{k})|\pm T_{2}|\mathbf{h}_{2}(\mathbf{k})|=n\pi,~n\in\mathbb{Z},\tag{4.b}\label{Con.b}
\addtocounter{equation}{1}
\end{align}
where $\mathcal{H}_{1,2}$ are expressed in terms of vectors $\mathbf{h}_{1,2}(\mathbf{k})$ according to Eq.~(\ref{HPauli}), and $\hat{\mathbf{h}}_j(\mathbf{k})=\mathbf{h}_j(\mathbf{k})/|\mathbf{h}_j(\mathbf{k})|$ (see Appendix \ref{appbandt}). The conditions given by Eqs.~(\ref{Con.a}) and (\ref{Con.b}) offer us a basic guideline to design our periodic quenching parameters to generate various topological states at will. (ii) Those $\mathbf{k}$ satisfying Eq.~(\ref{Con.a}) are called below ``prospective touching points" (PTPs). Substituting $\mathbf{h}_{1}(\mathbf{k})$ and $\mathbf{h}_{2}(\mathbf{k})$ at an arbitrary PTP into Eq. (\ref{Con.b}), one arrives at equations for quenching durations $T_{1}$ and $T_{2}$ in order to induce band touching.  Such equations define straight lines on a phase diagram in terms of $T_1$ and $T_2$.  These straight lines, tunable by choosing different integer $n$, are called band-touching lines. (iii) If $n$ is even (odd), then the band touching occurs at the quasienergy $0$ ($\pm\pi/T$). The band touching at $\pm\pi/T$ is unique in periodically driven systems.  (iv) As shown in Appendix \ref{appchira}, except for a special case, the band touching identified above leads to topological phase transitions. Thus, the band-touching lines depicted by Eq.~(\ref{Con.b}) are in general boundaries between different topological phases. (v) The change of the Floquet-band Chern numbers $\mathcal{C}$ across a topological phase transition line can be linked with the total chirality of the associated band-touching points \cite{PhysRevB.87.115402}.  It can be proven that the jump in $\mathcal{C}$ across a phase boundary is always the same along the entire phase transition line and  assumes an opposite sign  if the parity of $n$ is changed for the same PTP (see Appendix \ref{appchira}).

\section{Model results}\label{Results}
As an example to confirm our above analytical results, we consider a (generalized) Haldane model \cite{PhysRevB.87.115402,PhysRevB.83.115404}, but now periodically quenched. In the static case the model has topologically nontrivial bands, but with $|\mathcal{C}|=2$ at most with the third-neighbor (N3) hopping included. In the operator basis $C_{\mathbf{k}}=(c_{a\mathbf{k}},c_{b\mathbf{k}})^{T}$, its static Hamiltonian can be written as $H=\sum_{\mathbf{k}\in \text{BZ}}C_{\mathbf{k}}^{\dag}\mathcal{H}(\mathbf{k})C_{\mathbf{k}}$, where $\mathcal{H}(\mathbf{k})$ is in the form of Eq.~(\ref{HPauli}). The components of the associated $\mathbf{h}(\mathbf{k})$ are
\begin{align}\label{hamiltonian}
\begin{split}
&h_{x}(\mathbf{k})=t_{1}[1+\cos(\mathbf{k}\cdot\mathbf{a}_{1})+\cos(\mathbf{k}\cdot\mathbf{a}_{2})]+t_{3}\\
&~~~~~~~~~\times\{2\cos[\mathbf{k}\cdot(\mathbf{a}_1-\mathbf{a}_2)]+\cos[\mathbf{k}\cdot(\mathbf{a}_1+\mathbf{a}_2)]\},\\
&h_{y}(\mathbf{k})=t_{1}[\sin(\mathbf{k}\cdot\mathbf{a}_{1})+\sin(\mathbf{k}\cdot\mathbf{a}_{2})]\\
&~~~~~~~~~+t_{3}\sin[\mathbf{k}\cdot(\mathbf{a}_1+\mathbf{a}_2)],\\
&h_{z}(\mathbf{k})=2t_{2}\sin\phi\{\sin(\mathbf{k}\cdot\mathbf{a}_1)-\sin(\mathbf{k}\cdot\mathbf{a}_2)\\
&~~~~~~~~~-\sin[\mathbf{k}\cdot(\mathbf{a}_1-\mathbf{a}_2)]\}+M,
\end{split}
\end{align}
where $t_{j}$ are the $j$th-neighbor hopping amplitude, $\phi$ is the phase gained by $t_{2}$ hoppings, and $M$ is the mass term. The primitive vectors are $\mathbf{a}_1=(\sqrt{3}/2,3/2)$ and $\mathbf{a}_2=(-\sqrt{3}/2,3/2)$ in the unit of the hexagonal lattice constant. The model has point group $C_{6}$ ($M=0$) or $C_{3}$ ($M\neq0$) \cite{PhysRevLett.61.2015}. The $\varepsilon(\mathbf{k})$ term has been neglected because it does not affect the concerned topology here.

\begin{figure}
  \centering
  \includegraphics[scale=0.7]{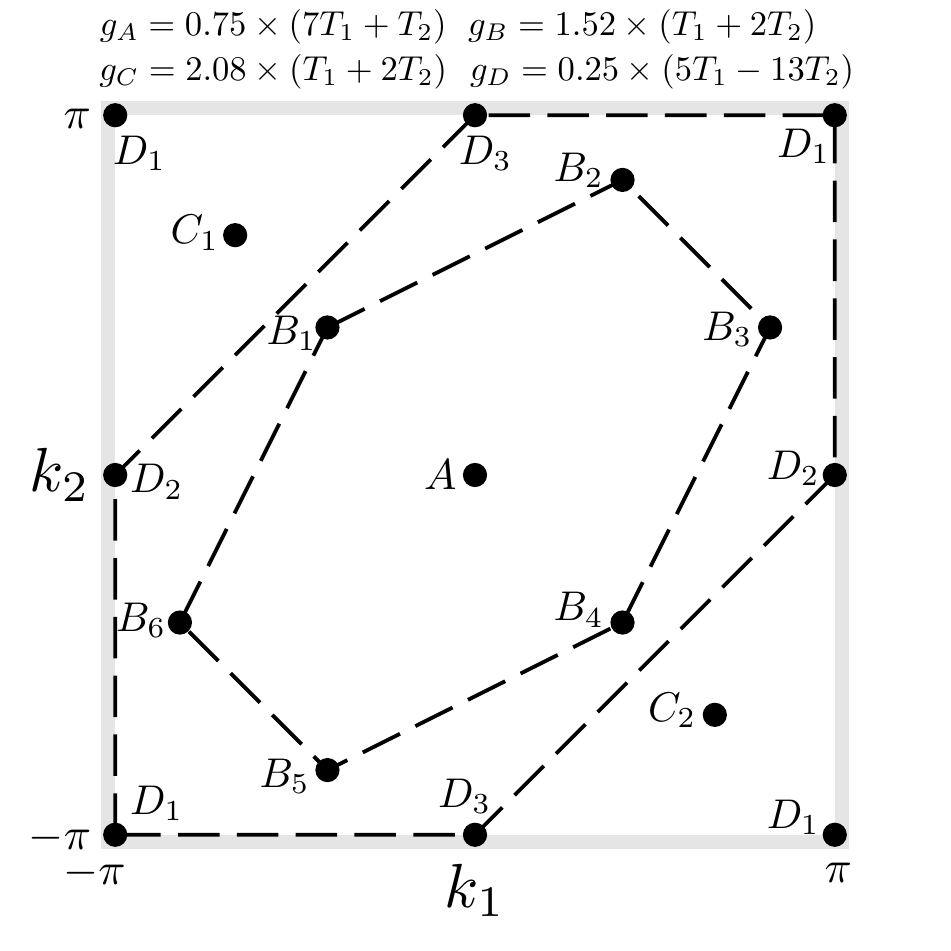}
  \caption{Examples of PTPs in BZ for a given set of ${\cal H}_{1}$ and ${\cal H}_{2}$ detailed in the main text, sorted into four classes $A$, $B$, $C$, and $D$ by their symmetry relation.  The coordinate bases of $k_{1},k_{2}$ are $(1/\sqrt{3},1/3),(-1/\sqrt{3},1/3)$, which are contravariant vectors of the primitive vectors $\bm{a}_{1},\bm{a}_{2}$. All plotted quantities in this work are dimensionless.}
  \label{starry}
\end{figure}

\begin{figure}
  \centering
  \includegraphics[trim=-6.88mm 0 0 0, scale=0.7]{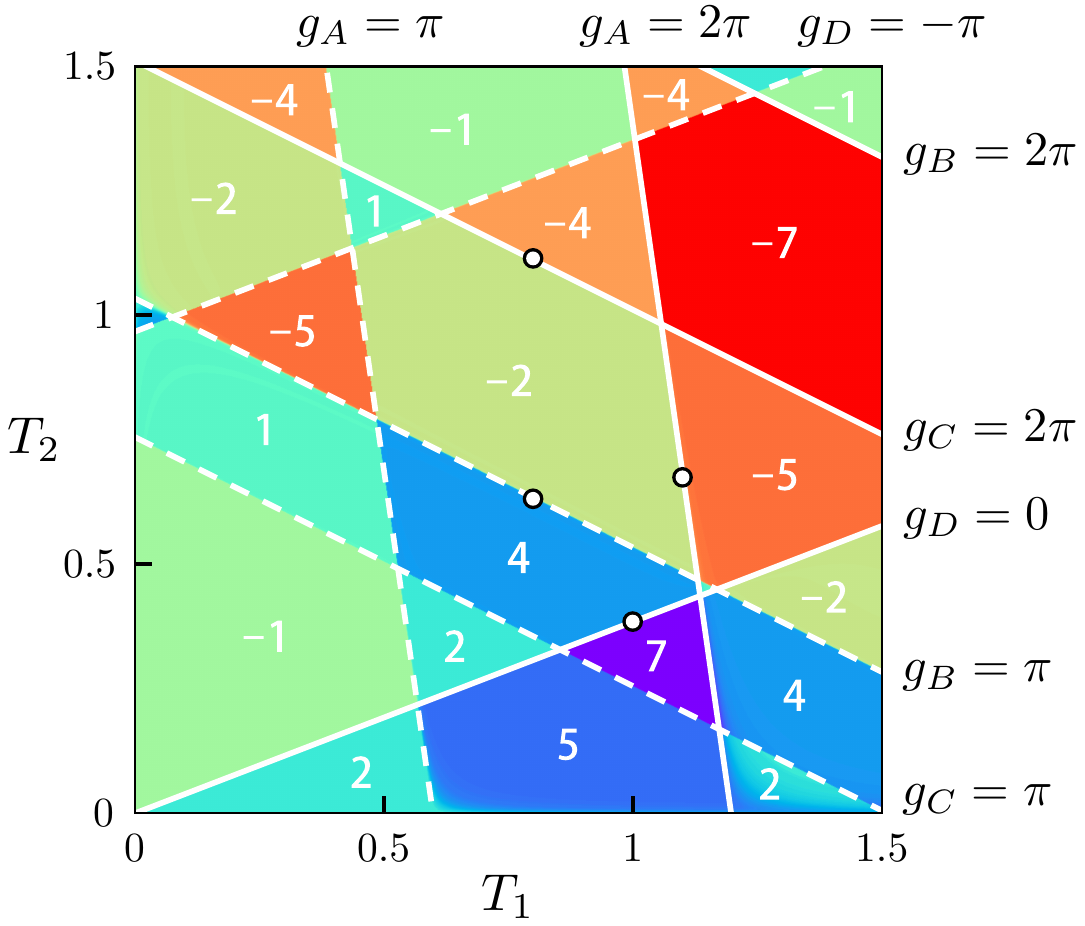}
  \caption{Topological phase diagram in the $T_1$-$T_2$ plane and the theoretically obtained band-touching lines (white lines), with different colors of the background representing the numerically obtained Chern numbers. The function forms of the straight phase transition lines are given on the right end of the figure panel, where $g_{A},g_{B},g_{C}$, and $g_{D}$ are specified in Fig.~\ref{starry}. Solid (dashed) lines are band-touching lines for even (odd) $n$.}
  \label{tangram}
\end{figure}

According to our general analysis, more PTPs lead to a richer phase diagram in terms of $T_1$ and $T_2$. We thus intend to choose $\mathbf{h}_{1}(\mathbf{k})$ and $\mathbf{h}_{2}(\mathbf{k})$ to generate more PTPs. We use $t_1$ to scale the energy unit. For illustration purpose, we fix $t_{2}=0.8$ and $M=0$ and periodically quench parameters $t_3$ and $\phi$. That is, we periodically set $t_{3}=0.75$ and $\phi=-\pi/6$ for duration $T_1$ and then let $t_{3}=-0.75$ and $\phi=-\pi/2$ for duration $T_2$. The changes in both $t_3$ and $\phi$ could be realized by modulating the optical lattice realizing the Haldane model \cite{PhysRevA.84.013607,PhysRevB.86.195129,Nature.13915}. Figure \ref{starry} shows the obtained PTPs. Since the $C_{6}$ symmetry of the Haldane model is maintained by the quenching, the PTPs can be classified according to their equivalence up to $C_{6}$ rotations. In particular, the PTPs in Fig.~\ref{starry} are sorted into four classes. Class $A$ has one member residing in the BZ center. Class $B$ has six members in the first BZ. Class $C$ has two PTPs because other equivalent partners go beyond the first BZ and return exactly to the first two PTPs if pulled back to the first BZ.  For a similar reason, class $D$ has only three independent PTPs on the edges of the first BZ. Due to $C_6$ symmetry,  all the PTPs in the same class  yield the same Eq.~(\ref{Con.b}).

For the above specified quenching protocol, the explicit expressions [denoted as $g_i$ ($i=A,~B,~C,~D$)] of the left hand side of Eq.~(\ref{Con.b}) are also shown in Fig.~\ref{starry} (see top) for each class. The band-touching lines are then given by $g_i=n\pi$. Different classes of PTPs and different values of $n$ then yield many band-touching and hence topological phase transition lines. In principle we can obtain as many band-touching lines as we wish. Figure \ref{tangram} shows a small portion of the topological phase diagram in the $T_1$-$T_2$ plane (that is, only the diagram for a small range of $T_1$ and $T_2$ is plotted). Different colors represent different Floquet-band Chern numbers. It is seen that the theoretical band-touching lines agree with the numerically obtained topological phase boundaries. Remarkably, topological phases with Floquet-band Chern numbers as large as $|\mathcal{C}|=7$ are obtained in our quenched Haldane system. Such large-Chern-number phases are not possible for the otherwise static Haldane model, unless much longer hopping terms are presented \cite{PhysRevB.87.115402}. Presumably, quenching system parameters in a system already experimentally realizable is much less challenging than fabricating very long hopping terms in a Hamiltonian. It indicates that besides inducing topologically nontrivial phases from topologically trivial one (see Ref. \cite{Nphys.1926} and Appendix \ref{Apptonon}), the periodic driving also supplies us a useful tool to achieve extremal topological states of matter absent in the static systems. We have also verified that the Chern number jumps between two topolgoical phases, denoted $\Delta \mathcal{C}$,  is indeed always a constant along the entire boundary. Furthermore, consistent with our general predictions, $\Delta\mathcal{C}$ for different parities of $n$ within the same class are indeed equal in magnitude but opposite in sign. For example, $\Delta \mathcal{C}$ along the line $g_{B}=\pi$ is $-6$; while it is $+6$ along the line $g_{B}=2\pi$.

\begin{figure*}
  \centering
  \begin{overpic}[width=0.9\textwidth]{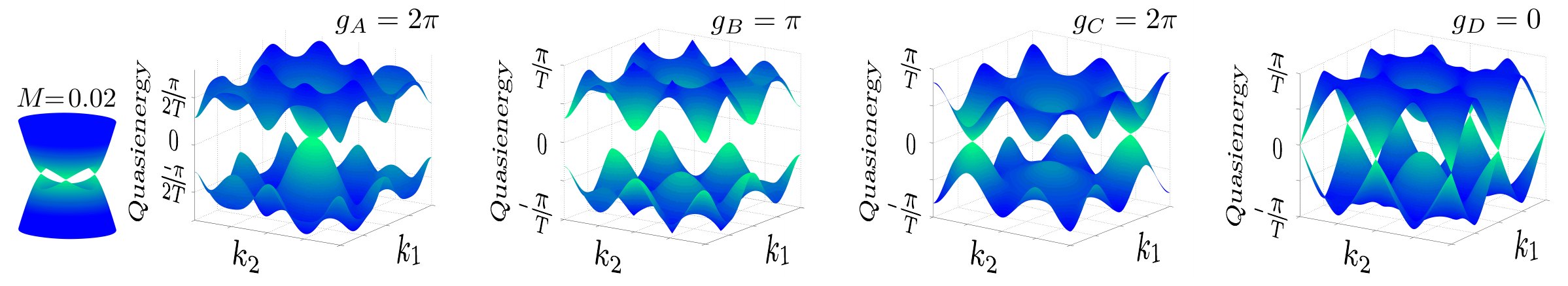}
  \put(1,17){\large (a)}
  \put(30,17){\large (b)}
  \put(54,17){\large (c)}
  \put(77,17){\large (d)}
  \end{overpic}
  \caption{Floquet band structure for the four band-touching cases marked in Fig.~\ref{tangram}. The parameters of (a)-(d) are $(T_{1},T_2)=(1.1,0.68)$, $(0.8,0.63)$, $(0.8,1.1)$, and $(1.0,0.38)$, with the band-touching lines given by  $g_{A}=2\pi$, $g_{B}=\pi$, $g_{C}=2\pi$, and $g_{D}=0$, respectively. The left subfigure of panel (a) shows the splitting of a nonlinear touching point into three Dirac points upon the addition of a perturbation to $M$. Note that the bands in panel (b) touch at $\pm \pi/T$.}
  \label{sea}
\end{figure*}

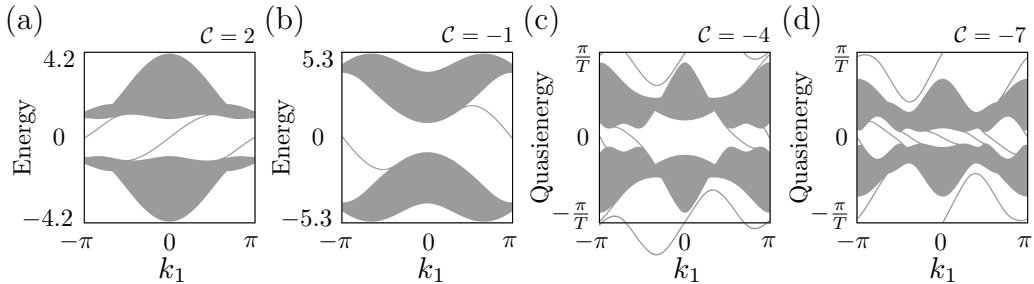
\begin{figure*}
  \centering
  \begin{overpic}[width=0.8\textwidth]{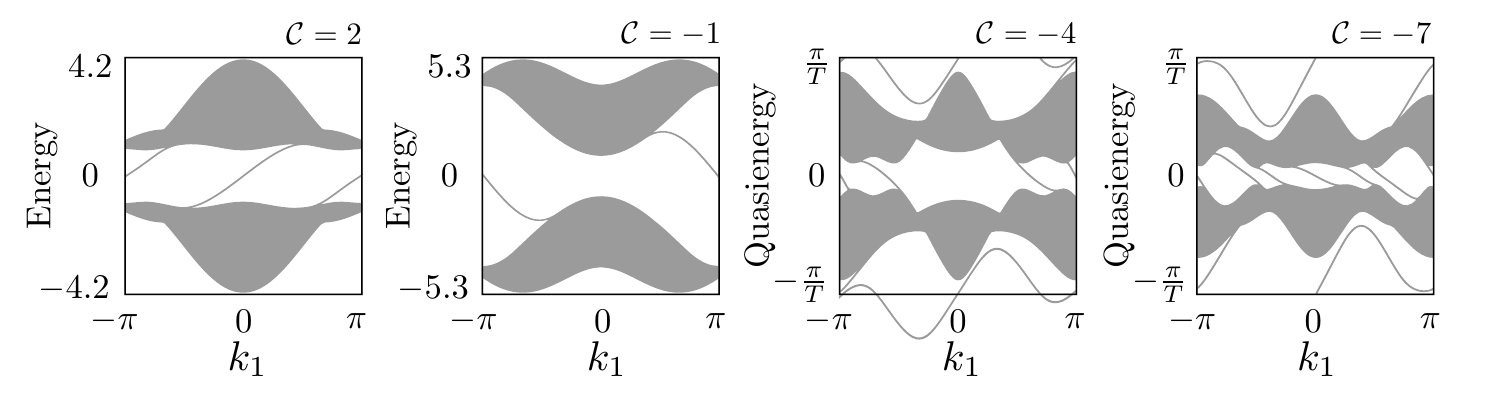}
   \put(1,25){\large (a)}
   \put(25,25){\large (b)}
   \put(49,25){\large (c)}
   \put(73,25){\large (d)}
  \end{overpic}
  \caption{Energy spectra of $H_{1}$ with $\mathcal{C}=2$ in (a) and $H_{2}$ with $\mathcal{C}=-1$ in (b). Floquet spectra of a system periodically quenched between $H_{1}$ and $H_{2}$, with $\mathcal{C}=-4$ and $(T_{1},T_{2})=(0.9,1.2)$ [with $\mathcal{C}=-7$ and $(T_1,T_2)=(1.3,1.2)$] in (c) [(d)].  }  \label{spectrum}
\end{figure*}

Figure \ref{sea} presents more detailed aspects of the Floquet bands,  for four choices of $T_1$ and $T_2$ associated with four classes of PTPs (already marked in Fig.~\ref{tangram}). The results confirm that the band touching occurs at $0$ [see Fig. \ref{sea}(a), (c) and (d)] and at $\pm\pi/T$ [see Fig. \ref{sea}(b)] when $n$ are even and odd, respectively. The positions of the band touching agree with what is observed in Fig.~\ref{starry}.

The jumps in the large Chern numbers across the phase boundaries shown in Fig.~\ref{tangram} are given by the total chirality of the band-touching points of a certain class (see Appendix \ref{appchira}). If one touching point has a linear energy dispersion, {\it i.e.,} a Dirac point,  then it has the chirality $\pm1$. Other touching points of the same class should have the same chirality. Consistent with this understanding, the total chirality for the band-touching cases shown in Fig.~\ref{sea}(b), (c), and (d), where the touching points are all Dirac points, are $-6$, $-2$, and $-3$. This hence identifies an underlying mechanism responsible for the generation of large-Chern-number phases. That is, the periodic quenching can induce many band-touching (Dirac) points of the same class, which then yields large Chern numbers. As a side result, the touching point in Fig. \ref{sea}(a) possesses a nonlinear dispersion, which may be understood as a limiting case with multiple Dirac points merging together. The topological number of such diabolical point equals to the total chirality of the hidden multiple Dirac points \cite{Springer.2013, PhysRevB.88.245422}. In the left panel of Fig.~\ref{sea}(a), this diabolical touching point is seen
to split into three Dirac points upon the addition of a perturbation to $M$. Thus $\Delta\mathcal{C}$ across the phase transition lines of class $A$ are $\pm3$, a fact consistent with the actual Chern numbers.

Periodic quenching of the Haldane model has thus created intriguing topological phases with Chern numbers ranging from $-7$ to $7$. Neither $H_1$ nor $H_2$ has these phases. We next investigate the Floquet spectra under the boundary condition of a semi-infinite strip with an edge at $y = 0$ \cite{PhysRevB.90.115423, PhysRevA.91.043625}. For comparison, Fig.~\ref{spectrum} also presents the energy spectra of the static $H_1$ and $H_2$.  As expected from the bulk-edge correspondence principle, the Chern number of the lower band equals to the difference in the numbers of chiral edge modes above and below the band \cite{PhysRevX.3.031005}.  As observed from Fig.~\ref{spectrum}(a) and (b), for $H_1$ there are only two chiral edge modes intersecting with the zero energy axis with positive group velocity (PGV), whereas for $H_2$ there is only one such edge mode with negative group velocity (NGV). Their energy-band Chern numbers are hence $2$ and $-1$. Markedly different,  the chiral edge (Floquet) modes of our periodically quenched system can go beyond the top and reenter the bottom of the quasienergy BZ \cite{PhysRevX.3.031005}. The case in Fig. \ref{spectrum}(c) has four chiral edge modes, three of which is above and the fourth one is below the lower band. The first three intersect with the zero quasienergy axis with NGV, whereas the bottom one intersects  with the  $\pm\pi/T$ axis five times in total, three times with PGV and two times with NGV. This yields a Floquet-band Chern number $\mathcal{C}=-3-1=-4$,  in agreement with  the numerical results in Fig.~\ref{tangram}. Similar analysis applies to Fig. \ref{spectrum}(d), where six edge modes bridge the gap above the lower band with NGV and the seventh one bridges the gap below the lower band with PGV, yielding $\mathcal{C}=-7$. Note that although the edge states in the Floquet topological phases is time-periodic, they still keep localized at the edges during their evolution (see Appendix \ref{appevolut}). These results further confirm that the above-obtained phases with large Chern numbers do have important implications for their edge states.

\section{Conclusions}\label{Conclusions}
Periodic quenching is not only experimentally feasible but also convenient in developing theoretical insights.  We have considered a class of periodically quenched two-band systems and explicitly obtained topological phase transition boundaries.  For a periodically quenched (generalized) Haldane model, the underlying mechanism for periodic quenching to induce large-Chern-number phases is explained through an induction of many Dirac-type band-touching points in the Floquet spectra. This work thus establishes a class of easily solvable, but very rich systems useful for further understandings and applications of Floquet topological phases.

\section*{Acknowledgments}\label{Acknowledgements}
JHA thanks Qing-Jun Tong for helpful discussions. JHA is supported by the Program for New Century Excellent Talents in University and the National Natural Science Foundation of China (Grant No. 11175072 and No. 11474139).  JG is funded by the Singapore Ministry of Education Academic Research Fund Tier 1 (WBS No. R-144-000-353-112).

\appendix

\section{Conditions for band touching}\label{appbandt}
We derive here in details the conditions under which the two Floquet bands of $U_T(\mathbf{k})$ touch each other.

A $2\times 2$ matrix $X$ can always be written as
\begin{equation}\label{Pauli}
X=\varepsilon{I}_{2\times2}+\mathbf{r}\cdot{\pmb\sigma},
\end{equation}
where $\mathbf{r}$ is a three-dimensional vector. The Hermitian conjugate of $X$ is given by $X^{\dag}=\varepsilon^{*}I_{2\times2}+\mathbf{r}^{*}\cdot{\pmb\sigma}$. If $X$ is unitary, then $\varepsilon$ and $\mathbf{r}$ must satisfy
\begin{equation}\label{Unitary}
|\varepsilon|^2+|\mathbf{r}|^2=1.
\end{equation}
Note that it is straightforward to evaluate the exponential of $X$ \cite{Springer.2003}
\begin{equation}\label{Expm}
e^{iX}=e^{i\varepsilon}(\cos|\mathbf{r}|\ I_{2\times2}+i\sin|\mathbf{r}|\ \hat{\mathbf{r}}\cdot{\pmb\sigma}),
\end{equation}
where $\hat{\mathbf{r}}=\mathbf{r}/|\mathbf{r}|$. For $X_{j}=\varepsilon_{j}I_{2\times2}+\mathbf{r}_{j}\cdot{\pmb\sigma}~(j=1,~2)$, their product becomes
\begin{equation}\label{Product}
\begin{split}
X_{1}X_{2}=&(\varepsilon_{1}\varepsilon_{2}+\mathbf{r_{1}}\cdot\mathbf{r_{2}})I_{2\times2}+\\
&(\varepsilon_{1}\mathbf{r_{2}}+\varepsilon_{2}\mathbf{r_{1}}+i\mathbf{r_{1}}\times\mathbf{r_{2}})\cdot{\pmb\sigma}.
\end{split}
\end{equation}

With these preparation steps and using the definitions $\mathcal{H}_{j}(\mathbf{k})=\varepsilon_{j}I_{2\times2}+\mathbf{h}_{j}\cdot{\pmb\sigma}$ and \begin{equation}\label{Floquet}U_T(\mathbf{k})=e^{-i\mathcal{H}_{2}(\mathbf{k})T_{2}}e^{-i\mathcal{H}_{1}(\mathbf{k})T_{1}},
\end{equation}
one then arrives at
\begin{equation}\label{Floquet[U]}
U_T(\mathbf{k})=e^{-i(\varepsilon_{1}+\varepsilon_{2})}(\varepsilon_{U}I_{2\times2}+i\mathbf{r}_{U}\cdot{\pmb\sigma}),
\end{equation}
where
\begin{equation}\label{Varepsilon[U]}
\begin{split}
\varepsilon_{U}=&\cos|T_{1}\mathbf{h}_{1}(\mathbf{k})|\cos|T_{2}\mathbf{h}_{2}(\mathbf{k})|\\
&-\cos\omega\sin|T_{1}\mathbf{h}_{1}(\mathbf{k})|\sin|T_{2}\mathbf{h}_{2}(\mathbf{k})|
\end{split}
\end{equation}
with $\cos\omega=\hat{\mathbf{h}}_{1}\cdot\hat{\mathbf{h}}_{2}$
and
\begin{equation}\label{R[U]}
\begin{split}
\mathbf{r}_{U}=&\sin|T_{1}\mathbf{h}_{1}| \sin|T_{2}\mathbf{h}_{2}|\ \hat{\mathbf{h}}_{1}\times\hat{\mathbf{h}}_{2}\\ &-\cos|T_{1}\mathbf{h}_{1}|\sin|T_{2}\mathbf{h}_{2}|\ \hat{\mathbf{h}}_{2}\\ &-\cos|T_{2}\mathbf{h}_{2}|\sin|T_{1}\mathbf{h}_{1}|\ \hat{\mathbf{h}}_{1}.
\end{split}
\end{equation}
Both $\varepsilon_{U}$ and $\mathbf{r}_{U}$ are seen to be real. The unitarity of $U_T(\mathbf{k})$ leads to $|\varepsilon_{U}|^{2}+|\mathbf{r}_{U}|^{2}=1$.

For two eigenvalues of $U_T(\mathbf{k})$ to be degenerate, the $\mathbf{r}_{U}$ vector in Eq.~(\ref{Floquet[U]}) should vanish such that the $U_T(\mathbf{k})$ matrix is proportional to an identity matrix. Equivalently, because $|\varepsilon_{U}|^{2}+|\mathbf{r}_{U}|^{2}=1$, the band touching happens when $|\varepsilon_{U}|=1$. This understanding then yields the following two possibilities for band touching \cite{Comment.1}:

Case I: $\cos\omega=\pm1$. Under this condition one has $\varepsilon_{U}=\cos(T_{1}|\mathbf{h}_{1}|\pm T_{2}|\mathbf{h}_{2}|)$. Then degeneracy or band touching occurs at $\mathbf{k}$ if
\begin{equation}\label{Criteria}
\begin{split}
&\hat{\mathbf{h}}_{1}(\mathbf{k})=\pm\hat{\mathbf{h}}_{2}(\mathbf{k}),\\
&T_{1}|\mathbf{h}_{1}(\mathbf{k})|\pm T_{2}|\mathbf{h}_{2}(\mathbf{k})|=n\pi,~~~~~
\end{split}
\end{equation}
where $n\in\mathbb{Z}$. In particular, if $n$ is even, then it is evident that $\varepsilon_{U}=1$, yielding a zero quasienergy. If $n$ is odd, then $\varepsilon_{U}=-1$, yielding eigenvalue $-1$ and as such, the quasienergy of the eigenvalues of $U_T(\mathbf{k})$ is $e^{\pm i \pi}$. In the latter case, the bands touch at quasienergy $\pm \pi/T$ instead of 0.
  
Case II:   $\sin|T_{1}\mathbf{h}_{1}|\sin|T_{2}\mathbf{h}_{2}|=0$. In this case $\varepsilon_{U}= \cos|T_{1}\mathbf{h}_{1}| \cos|T_{2}\mathbf{h}_{2}|$. Then band touching occurs under the conditions
\begin{equation}\label{Criteria[B]}
\begin{split}
&T_{1}|\mathbf{h}_{1}(\mathbf{k})|=n_{1}\pi,~n_{1}\in\mathbb{Z};\\
&T_{2}|\mathbf{h}_{2}(\mathbf{k})|=n_{2}\pi,~n_{2}\in\mathbb{Z}.~~~~~~~~~
\end{split}
\end{equation}

However, the band-touching points determined by Eq. (\ref{Criteria[B]}) do not give rise to topological phase transitions in our periodic quenching protocol (details in the next section). For this reason, in the main text we focus on the band touching points determined by Eq.~(\ref{Criteria}).

Note in the passing that due to $|\varepsilon_{U}|^{2}+|\mathbf{r}_{U}|^{2}=1$, there always exists a $\mathbf{r}_{u}$ satisfying $\varepsilon_{U}=\cos|\mathbf{r}_{u}|$ and $\mathbf{r}_{U}=\hat{\mathbf{r}}_{u}\sin|\mathbf{r}_{u}|$.
Equation (\ref{Floquet[U]}) thus becomes
\begin{equation}\label{Floquet[u]}
U_T(\mathbf{k})=e^{-i(\varepsilon_{1}+\varepsilon_{2})}(\cos|\mathbf{r}_{u}|\ I_{2\times2}+i\sin|\mathbf{r}_{u}|\ \hat{\mathbf{r}}_{u}\cdot{\pmb\sigma}).
\end{equation}
Comparing this form of $U_T(\mathbf{k})$ with Eq.~(\ref{Expm}), one obtains
\begin{equation}\label{agggd}
\mathcal{H}_{\text{eff}}(\mathbf{k})
=\frac{1}{T_1+T_2}[(\varepsilon_{1}+\varepsilon_{2})I_{2\times2}-\mathbf{r}_{u}\cdot{\pmb\sigma}]
\end{equation}
The associated $\mathbf{h}_{\text{eff}}$ vector [see Eq.~(\ref{HPauli})] for this Hamiltonian is then given by \cite{Comment.2}
\begin{equation}\label{Heff[U]}
\mathbf{h}_{\text{eff}}=-\frac{1}{T_1+T_2}\arccos(\varepsilon_{U})\hat{\mathbf{r}}_{U}.
\end{equation}
The Chern number of the lower Floquet band can then be calculated from
\begin{equation}\label{Chern}
\mathcal{C}=\frac{1}{4\pi}\int_\text{BZ}\d^{2}\mathbf{k}~ \frac{\mathbf{h}_{\text{eff}}}{|\mathbf{h}_{\text{eff}}|^{3}}\cdot(\partial_{k_{x}}\mathbf{h}_{\text{eff}}\times\partial_{k_{y}}\mathbf{h}_{\text{eff}}).
\end{equation}

\section{Chirality of touching points}\label{appchira}
The band touching conditions depicted by (\ref{Criteria}) and (\ref{Criteria[B]}) offer necessary conditions for having a topological phase transition. Here we derive two rules regarding the potential jumps in the Chern number (of one Floquet band) across a band-touching line in the $T_1$-$T_2$ plane.

The jump in the Floquet-band Chern number is given by
\begin{equation}\label{Sum}\Delta \mathcal{C}=\sum_{i}\chi_i,\end{equation}
where $\chi_{i}=\sgn[(\partial_{k_{x}}\mathbf{h}_{\text{eff}}\times\partial_{k_{y}}\mathbf{h}_{\text{eff}})\cdot\delta\mathbf{h}_{\text{eff}}]_{\mathbf{k}=\mathbf{k}_{i}}$ defines the chirality of the $i$th touching point $\mathbf{k}_{i}$, and $\delta\mathbf{h}_{\text{eff}}$ is the change in $\mathbf{h}_{\text{eff}}$ across a touching point. Note that Eq.~(\ref{Sum}) is valid only when $\partial_{k_{x}}\mathbf{h}_{\text{eff}}\times\partial_{k_{y}}\mathbf{h}_{\text{eff}}$ is nonzero \cite{PhysRevB.85.165456}. This requires that the dispersion relation at the touching point is linear. That is, the touching point is of the Dirac type. By contrast, if a touching point has a nonlinear dispersion relation, then $\partial_{k_{x}}\mathbf{h}_{\text{eff}}\times\partial_{k_{y}}\mathbf{h}_{\text{eff}}$ is zero \cite{PhysRevB.85.165456, PhysRevB.87.115402, PhysRevB.83.245125}. This may be understood as a merging of multiple Dirac points \cite{PhysRevB.87.115402,Springer.2013}. Such a merging can be split again by adding a perturbation [see Fig.~\ref{sea}(a)].

Using the expressions of $\mathbf{h}_{\text{eff}}$ in the preceding section,  the chirality $\chi_i$ can also be calculated from
\begin{equation}\label{Chi[r]}
\chi_{i}=-\sgn[(\partial_{k_{x}}\mathbf{r}_{U}\times \partial_{k_{y}}\mathbf{r}_{U})\cdot\delta\mathbf{r}_{U}]_{\mathbf{k}=\mathbf{k}_{i}},
\end{equation}
where $\delta\mathbf{r}_{U}$ is the change in $\mathbf{r}_{U}$ across a band touching point. Based on this, we derive below the explicit form of $\Delta \mathcal{C}$ for the two band-touching cases identified in the preceding section.

\subsection{Case I}
For the touching points determined by Eq. (\ref{Criteria}), $\mathcal{H}_{1}(\mathbf{k})$ and $\mathcal{H}_{2}(\mathbf{k})$ commute with each other. Then from the Baker-Campbell-Hausdorff formula \cite{Springer.2003}, Eq.~(\ref{Floquet}) becomes
\begin{equation}\label{Floquet[B]}
U_T(\mathbf{k})=e^{-i[\mathcal{H}_{2}(\mathbf{k})T_{2}+\mathcal{H}_{1}(\mathbf{k})T_{1}]}.
\end{equation}
Consider variations of $T_{1}$ and $T_{2}$ across a band touching line determined by $T_{1}|\mathbf{h}_{1}|\pm T_{2}|\mathbf{h}_{2}|=n\pi$, i.e., $T_{1}|\mathbf{h}_{1}|\pm T_{2}|\mathbf{h}_{2}|$ changes from $n\pi-\delta{t}$ to $n\pi+\delta{t}$. Using Eqs. (\ref{Expm}) and (\ref{Floquet[B]}), one can prove that $U_T(\mathbf{k})$, to the first order of $\delta{t}$, changes from $I_{2\times2}\pm i\,\delta{t}\,\hat{\mathbf{h}}_{1}\!\cdot{\pmb\sigma}$ to $I_{2\times2}\mp i\,\delta{t}\,\hat{\mathbf{h}}_{1}\!\cdot{\pmb\sigma}$, where the upper (lower) signs corresponds to $n$ being even (odd) numbers. This yields
\begin{equation}\label{Delta[ru]}
\delta\mathbf{r}_{U}=2(-1)^{n-1}\delta{t}\,\hat{\mathbf{h}}_{1}.
\end{equation}
We can see from Eq. (\ref{Chi[r]}) that only the component of $\partial_{k_{x}}\mathbf{r}_{U}\times\partial_{k_{y}}\mathbf{r}_{U}$ parallel to $\delta\mathbf{r}_{U}$ contributes to $\chi_i$. Using Eqs. (\ref{Criteria}), the parallel component can be written in the following compact forms, {\it i.e.},
\begin{equation}\notag
(\partial_{k_{x}}\mathbf{r}_{U}\times \partial_{k_{y}}\mathbf{r}_{U})_{\parallel}= \sin^{2}\theta\ (\bm{\mathcal{F}}_{k_x,1}-\bm{\mathcal{F}}_{k_x,2})\times (\bm{\mathcal{F}}_{k_y,1}-\bm{\mathcal{F}}_{k_y,2}),
\end{equation}
if $\hat{\mathbf{h}}_1(\mathbf{k})=\hat{\mathbf{h}}_2(\mathbf{k})$ and
\begin{equation}\notag
(\partial_{k_{x}}\mathbf{r}_{U}\times \partial_{k_{y}}\mathbf{r}_{U})_{\parallel}= \sin^{2}\theta\ (\bm{\mathcal{F}}_{k_x,1}+\bm{\mathcal{F}}_{k_x,2})\times (\bm{\mathcal{F}}_{k_y,1}+\bm{\mathcal{F}}_{k_y,2}),
\end{equation}
if $\hat{\mathbf{h}}_1(\mathbf{k})=-\hat{\mathbf{h}}_2(\mathbf{k})$, where $\theta=-|T_{1}\mathbf{h}_{1}|$ and
\begin{equation}\label{F[AlphaBeta]}
\bm{\mathcal{F}}_{\alpha,j}=\sin{\theta}\ \hat{\mathbf{h}}_{1}\times\partial_{\alpha}\hat{\mathbf{h}}_{j}+\cos{\theta}\ \partial_{\alpha}\hat{\mathbf{h}}_{j},
\end{equation}
with $\alpha=k_{x},~k_{y}$ and $j=1,~2$.
Based on the fact that $\partial_{\alpha}\hat{\mathbf{h}}_{j}$, $\hat{\mathbf{h}}_{1}\times\partial_{\alpha}\hat{\mathbf{h}}_{j}$, and $\hat{\mathbf{h}}_{1}$ are perpendicular each other, and $|\partial_{\alpha}\hat{\mathbf{h}}_{j}|=|\hat{\mathbf{h}}_{1}\times\partial_{\alpha}\hat{\mathbf{h}}_{j}|$, one can obtain $|\bm{\mathcal{F}}_{\alpha,j}|=|\partial_{\alpha}\hat{\mathbf{h}}_{j}|$.
This makes it clear that $\bm{\mathcal{F}}_{\alpha,j}$ lies in a plane perpendicular to $\hat{\mathbf{h}}_{1}$, and the angle $\theta$ becomes the angel between vectors $\bm{\mathcal{F}}_{\alpha,j}$ and $\partial_{\alpha}\hat{\mathbf{h}}_{j}$. Given $\hat{\mathbf{h}}_{1}$ and $\partial_{\alpha}\hat{\mathbf{h}}_{j}$, a change in $\theta$ only rotates $\bm{\mathcal{F}}_{\alpha,j}$ around the $\hat{\mathbf{h}}_{1}$ axis, but
maintaining the relative angels between vectors $\bm{\mathcal{F}}_{\alpha,1}$ and $\bm{\mathcal{F}}_{\alpha,2}$. It then follows that $(\bm{\mathcal{F}}_{k_x,1}\mp\bm{\mathcal{F}}_{k_x,2})\times (\bm{\mathcal{F}}_{k_y,1}\mp\bm{\mathcal{F}}_{k_y,2})$ is independent of $\theta$.

Substituting $(\partial_{k_{x}}\mathbf{r}_{U}\times \partial_{k_{y}}\mathbf{r}_{U})_{\parallel}$ and Eq. (\ref{Delta[ru]}) into Eq. (\ref{Chi[r]}), one may further reduce the expression for the chirality of the $i$th band-touching point, obtaining
\begin{equation}\label{Chi[F]}
\notag
\chi_{i}=(-1)^n\sgn[(\bm{\mathcal{F}}_{k_x,1}\mp\bm{\mathcal{F}}_{k_x,2})\times (\bm{\mathcal{F}}_{k_y,1}\mp\bm{\mathcal{F}}_{k_y,2}) \cdot \hat{\mathbf{h}}_{1}]_{\mathbf{k}=\mathbf{k}_{i}},
\end{equation}
where the factors $\sin^{2}\theta$ and $2\delta{t}$ are dropped because they do not contribute to the sign function and the notation $\mp$ has the same correspondence to the one in Eqs. (\ref{Criteria}).  If $\mathbf{h}_{1}$ is normal to $(\bm{\mathcal{F}}_{k_x,1}\mp\bm{\mathcal{F}}_{k_x,2})\times (\bm{\mathcal{F}}_{k_y,1}\mp\bm{\mathcal{F}}_{k_y,2})$, then the above expression indicates that the chirality would be zero. In this special case, the band touching would not lead to a jump in the Floquet-band Chern numbers. In general situations, there would be a nonzero jump that signals a topological phase transition.  Remarkably, because $(\bm{\mathcal{F}}_{k_x,1}\mp\bm{\mathcal{F}}_{k_x,2})\times (\bm{\mathcal{F}}_{k_y,1}\mp\bm{\mathcal{F}}_{k_y,2})$ is shown to be independent to $\theta$ or $T_1$ (and equivalently independent of $T_2$), along one whole band-touching line, the chirality
value would be the same and thus the jump in the associated Chern number would be the same along the entire band-touching line. In addition, the chirality for  band-touching lines indexed by different parities of $n$ is only differed by their sign. This type of interlaced phase transition lines with opposite chirality values prevents the band Chern numbers from growing violently.
The hope in achieving large Chern numbers is then mainly with an increase of equivalent touching points in the first BZ (that is, due to the emergence of multiple Dirac points) such that the sum in Eq.~(\ref{Sum}) can still yield large values.

\begin{figure*}
  \centering
  \includegraphics[width=0.6\textwidth]{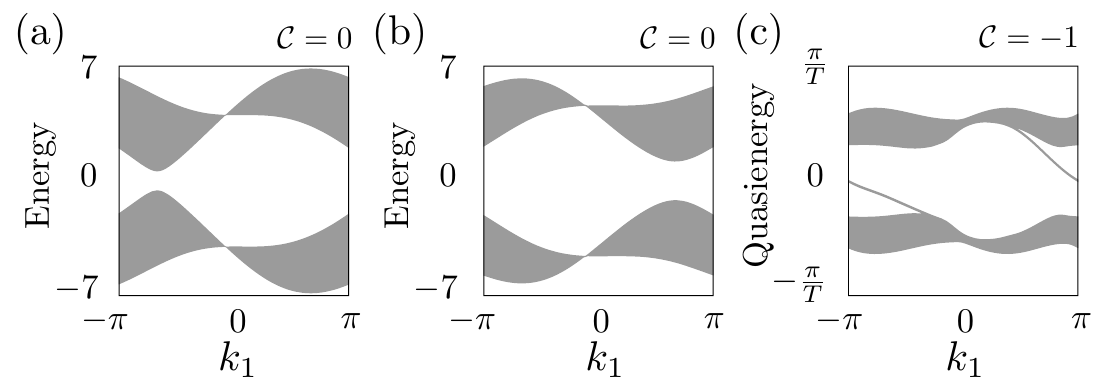}
  \caption{Energy spectra of $H_{1}^{\prime}$ in (a) and $H_{2}^{\prime}$ in (b) with $\mathcal{C}=0$. (c): Floquet spectrum with $\mathcal{C}=-1$ when the system is periodically quenched between $ H_{1}^{\prime}$ and $H_{2}^{\prime}$  with $(T_{1},T_{2})=(0.9,0.8)$. Other parameters are described in the text.}
  \label{Tree}
\end{figure*}

\begin{figure*}
  \centering
  \includegraphics[width=0.99\textwidth]{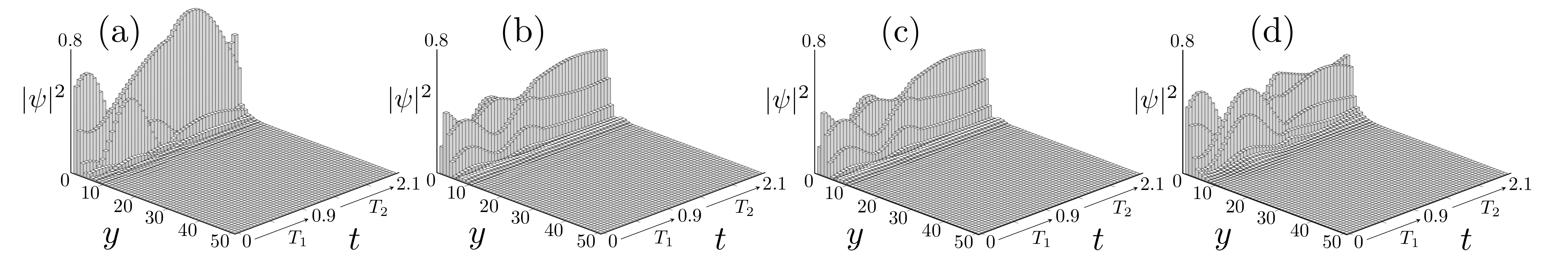}
  \caption{Evolution of the occupation probabilities of four Floquet edge states within one period, all chosen from Fig.~4(c) in the main text. Edge states in (a)-(c) intersect with the zero quasienergy axis, and edge state in (d) is the one interacting with the $\pm\pi/T$  quasienergy axis at $k_{x}=0$. }
  \label{Cube}
\end{figure*}

\subsection{Case II}
Next we discuss the band touching case under the condition of Eq.~(\ref{Criteria[B]}). Suppose $T_{j}|\mathbf{h}_{j}|$ changes from $n_{j}\pi-\delta{t_{j}}/2$ to $n_{j}\pi+\delta{t_{j}}/2$, then a calculation to the first order of $\delta{t_{j}}$ yields
\begin{equation}\label{Delta[ruii]}
\delta\mathbf{r}_{U}=(-1)^{1+n_{1}+n_{2}}(\delta{t_{1}}\,\hat{\mathbf{h}}_{1} +\delta{t_{2}}\,\hat{\mathbf{h}}_{2}).
\end{equation}
From Eqs.~(\ref{Criteria[B]}) and (\ref{R[U]}), one has
\begin{equation}\label{Partial[ruii]}
\partial_{\alpha}\mathbf{r}_{U}=(-1)^{1+n_{1}+n_{2}}(\partial_{\alpha}|T_{2}\mathbf{h}_{2}|\ \hat{\mathbf{h}}_{2} +\partial_{\alpha}|T_{1}\mathbf{h}_{1}|\ \hat{\mathbf{h}}_{1}),
\end{equation}
where $\alpha=k_{x},~k_{y}$. Equations (\ref{Delta[ruii]}) and (\ref{Partial[ruii]}) show that the three vectors $\delta\mathbf{r}_{U}$, $\partial_{k_{x}}\mathbf{r}_{U}$, $\partial_{k_{y}}\mathbf{r}_{U}$ are all on the plane spanned by vectors $\hat{\mathbf{h}}_{1}$ and $\hat{\mathbf{h}}_{2}$. Hence, the triple product of these vectors in Eq. (\ref{Chi[r]}) must be zero. Evidently then,   all band-touching points in this case have zero charility
 and will not induce a change in the Chern numbers.

\section{Turning a topologically trivial system to a topologically nontrivial one} \label{Apptonon}
 For completeness, we also note that our periodic quenching protocol can turn a topologically trivial system into a nontrivial one. Figure~\ref{Tree} shows one such example using again the periodically quenched Haldane model described in the main text, also under the boundary condition of a semi-infinite strip with an edge at $y = 0$.
 The static parameters are $t_{2}=0.6$ and $M=-3.7$, and the periodic quenching is applied such that $t_{3}=-0.5$ and $\phi=-0.5\pi$ for $H_{1}^{\prime}$, and $t_{3}=-0.1$ and $\phi=0.3\pi$ for $H_{2}^{\prime}$.  It is seen that one edge mode in the Floquet spectrum is formed in
 Fig.~\ref{Tree}(c), while the original static Hamiltonians $H^{\prime}_{1}$ and $H^{\prime}_{2}$ have no edge mode present.

\section{Aspects of edge state evolution}\label{appevolut}
 What is considered in the main text is based on Floquet eigenstates and only the outcome of time evolution at integer multiples of $T=T_1+T_2$ is studied. It is of interest to
 check the dynamical behavior of Floquet edge states within one quenching period.
 We present in Fig.~\ref{Cube} the evolution of four such Floquet edge states, all chosen from
 Fig.~2(c) in the main text. It is seen (i) that after a whole period of evolution all the edge states return to their initial states, and that (ii) at all times within one period these states remain well localized on the edge.  This further confirms the localization behavior of Floquet edge states.

\bibliographystyle{apsrev4-1}
\bibliography{main-resub-v3}

\begin{thebibliography}{60}
\bibitem [{Com({\natexlab{a}})}]{Comment.1}%
  \BibitemOpen
  \href@noop {} {}$|\varepsilon_{U}| \!\leq\!
  \big|\cos|T_{1}\mathbf{h}_{1}|\cos|T_{2}\mathbf{h}_{2}|\big| \!+\!
  \big|\cos\omega\sin|T_{1}\mathbf{h}_{1}|\sin|T_{2}\mathbf{h}_{2}|\big|
  \!\leq\! \big|\cos|T_{1}\mathbf{h}_{1}|\cos|T_{2}\mathbf{h}_{2}|\big| \!+\!
  \big|\sin|T_{1}\mathbf{h}_{1}|\sin|T_{2}\mathbf{h}_{2}|\big| \!\leq\! 1$. The
  second equality requires
  $\big|\cos\omega\sin|T_{1}\mathbf{h}_{1}|\sin|T_{2}\mathbf{h}_{2}|\big| \!=\!
  \big|\sin|T_{1}\mathbf{h}_{1}|\sin|T_{2}\mathbf{h}_{2}|\big|$, which means
  $\sin|T_{1}\mathbf{h}_{1}|\sin|T_{2}\mathbf{h}_{2}|=0$ or
  $|\cos\omega|=1$.\BibitemShut {Stop}%
\bibitem [{Com({\natexlab{b}})}]{Comment.2}%
  \BibitemOpen
  \href@noop {} {}With no contribution to Chern numbers, the identity matrix
  term in Eq. (\ref{agggd}) is generally neglected. One can then yield
  $|\mathbf{r}_{u}|\in[0,\pi]$ from the convention that quasienergies are
  chosen to lie in $(-\pi/T,\pi/T]$.\BibitemShut {Stop}%
\end{thebibliography}

\end{document}